\documentstyle[psfig]{mn}

\def\etal{{\it et al. }}
\title[Globular Clusters in NGC 1052
and NGC 7332]
{Keck Imaging of the Globular Cluster Systems in the Early--type 
Galaxies NGC 1052 and NGC 7332}

\author[Forbes, Georgakakis, \& Brodie] {
  Duncan A. Forbes$^{1}$\thanks{dforbes@swin.edu.au}, 
  Antonis E. Georgakakis$^{2}$\thanks{age@star.sr.bham.ac.uk},
  Jean P. Brodie$^{3}$\thanks{brodie@ucolick.org}\\ \\
  $^1$ Astrophysics \& Supercomputing, Swinburne University,
  Hawthorn, VIC 3122, Australia\\
  $^2$ School of Physics and Astronomy, University of Birmingham,
  Edgbaston, B15 2TT, UK\\
  $^3$ Lick Observatory, University of California,
 Santa Cruz, CA 95064, USA\\
}
\begin{document}
\maketitle

\begin{abstract}

The presence of two globular cluster subpopulations in
early--type galaxies is now the norm rather than the
exception. Here we present two more examples for which the host
galaxy appears to have undergone a recent merger.
Using multi--colour Keck imaging of NGC 1052 and NGC 7332 we find
evidence for a bimodal globular cluster colour distribution in
both galaxies, 
with roughly equal numbers of blue and red globular clusters. The
blue ones have similar colours to those in the Milky Way halo and
are thus probably very old and metal--poor. 
If the red GC subpopulations are  
at least solar metallicity, then stellar population models
indicate young ages. We discuss the origin of globular clusters
within the framework of formation models. We conclude that recent
merger events in these two galaxies have had little effect on their
overall GC systems.
We also derive globular cluster 
density profiles, global specific frequencies and in
the case of NGC 1052, radial colour gradients and azimuthal
distribution. In general these globular cluster properties are normal for
early--type galaxies.

\end{abstract}


\section{Introduction}\label{sec_intro}

Perhaps the most noteworthy of recent findings in extragalactic globular
cluster (GC) research
is the discovery of bimodal color distributions in the GC systems of
massive elliptical galaxies. Bimodality is taken to
 indicate the presence of 
two distinct GC subpopulations.

There are at least three scenarios for producing bimodal color
distributions in the GC systems of elliptical galaxies. {\it i)} The merger
of two gas--rich (spiral) galaxies may lead to the formation of an
elliptical galaxy and create an additional population of GCs in
the process
(Schweizer 1987, Ashman \& Zepf 1992). Since the population 
produced in the
merger formed from enriched gas it should be of higher metallicity and
redder than the indigenous population. Here the expectation is
that the age
of the red clusters should be similar to the epoch of the merger
event. Note that, according to Worthey (1994)
stellar population models, this
second metal--rich population will always be redder in (U--V) than the old
metal--poor population if it is older than $\sim$1 Gyr but it will
be redder in (B--I) only if it is older than $\sim$2 Gyr.  {\it ii)} A
multi--phase collapse (Forbes, Brodie \& Grillmair 1997) 
can also produce a bimodal color distribution with the blue clusters
forming in an early chaotic phase of galaxy formation from metal-poor gas
and the red clusters forming a few Gyr later from enriched gas in the same
phase that produces the bulk of the galaxy starlight. {\it iii)} Cot\'e,
Marzke \& West (1998) describe the build-up of the GC systems of bright
ellipticals via the accretion of mostly metal-poor GCs from dwarf
galaxies
In this third picture the red GCs are indigenous and the blue
ones acquired. These three scenarios predict a different
separation in age between the red and blue GCs. The merger model predicts
an old population ($\sim$age of the universe) plus a young
population with the age of the merger.  A 
multi--phase collapse also has an old population and 
one slightly younger ($\sim$2--4 Gyr younger) than the other.
The accretion scenario predicts that the age distribution of the 
blue and the red GCs will be similar unless a recent 
merger product has been
accreted.

The key to discriminating between these possibilities is to
 determine ages and
metallicities for both the blue and red GC subpopulations.
In principle spectroscopy is the best way to determine age and
 metallicity 
independently, however this requires very high signal-to-noise data which 
is rarely available for extragalactic GCs. Alternatively one can
 use the difference in colour of the two subpopulations, from 
photometry, to {\it constrain} the age. 
The latter method has only been applied to a
 small number 
of galaxies (Whitmore et al. 1997; Puzia et al. 1999; Brown \etal
2000; Georgakakis et al. 2000). 

Here we present data from the same Keck observing run as Georgakakis 
et al. (2000) for two additional galaxies -- NGC 
1052 and NGC 7332. Both galaxies reveal signs of a recent merger
or interaction, and are therefore ideal candidates to search for
young GCs. 
Details of the data reduction, 
GC selection process and contamination issues can be found in the initial 
paper (on the young elliptical galaxy NGC 6702 and its two GC
 subpopulations).

In section \ref{sec_obs} the observations and data reduction 
techniques are 
briefly described. Section \ref{galaxy} presents basic results for the 
host galaxies. Section \ref{sec_sel} 
describes the GC selection, 
with results given in section \ref{sec_res}. A 
discussion is given in \ref{sec_disc}. 
Throughout this  paper we adopt $H_0=75\,\mathrm{km\,s^{-1}\,Mpc^{-1}}$.  

\section{Observations and data reduction}\label{sec_obs}

Broad--band imaging of NGC 1052 and NGC 7332, 
in the  B, V and I, filters was
carried out at the Keck-II telescope on 1999 August 17th, 
using the the Low
Resolution Imaging Spectrometer (LRIS; Oke et al. 1995). The LRIS
instrument, equipped with a TEK $2048\times2048$ CCD, is mounted on the
Cassegrain focus providing a
$0.215\,\mathrm{arcsec\,pixel^{-1}}$ imaging scale and a
$6^{\prime}\times8^{\prime}$ field--of--view. The total exposure
 times and seeing 
conditions are given in Table 1. 

\begin{table*} 
\footnotesize 
\begin{center} 
\begin{tabular}{lccccc} 
\hline 
   Name  & B & V & I & Seeing\\ 
          &  (secs)    &  (secs) & (secs) & (arcsec)\\
\hline
NGC~1052 & 2400 & 1200 & 300 & 0.8\\
NGC~7332 & 3600 & 1800 & 1200 & 0.8\\

\hline
\end{tabular} 
\end{center} 
\caption{Observations. The total exposure time and average seeing
conditions are given.}\label{tab1}
\normalsize  
\end{table*}

The data were reduced following standard procedures, using IRAF
software. 
The reduced images were found to be flat to better than $\sim2\%$. 
Photometric calibration was performed using standard stars
from Landolt (1992). The photometric accuracy, estimated  using these 
standard stars, is $\pm0.02$\,mag in all three bands.

\section{Galaxy properties}\label{galaxy}

In Figures \ref{n1052_grey} and
\ref{n7332_grey} we show the 
B band Keck images of NGC 1052 and NGC 7332 respectively. 
Both images reveal numerous unresolved objects. 
Table 2 lists some basic properties of the galaxies. 

\begin{figure} 
\caption
{$V$-band image of NGC 1052 covering
$\approx5.6^{\prime}\times7.0^{\prime}$. North is up and East is
left. {\it Please email dforbes@swin.edu.au for a copy of this
figure}}
\label{n1052_grey}  
\end{figure}

\begin{figure} 
\caption
{$V$-band image of NGC 7332 covering  
$\approx5.6^{\prime}\times7.0^{\prime}$. North is
up and East is left. {\it Please email dforbes@swin.edu.au for a copy of this
figure}}
\label{n7332_grey}
\end{figure}

\begin{table*} 
\footnotesize 
\begin{center} 
\begin{tabular}{lcccccc} 
\hline 
   Name  &  Type & A$_B$ & Distance & M$_B$ & B--V & B$_{TO}$ \\ 
          &  (RC3)    &  (mag) & (Mpc) & (mag) & (mag) & (mag) \\
\hline
NGC~1052 & E4 & 0.114 & 17.70 & --19.27 & 0.97 & 24.74\\
NGC~7332 & S0pec & 0.161 & 15.28 & --19.06 & 0.88 & 24.42\\

\hline
\end{tabular} 
\end{center} 
\caption{Galaxy properties. 
The extinction A$_B$ comes from Schlegel et al. (1998), the 
distance includes a Virgocentric infall correction for 
$H_0=75\,\mathrm{km\,s^{-1}\,Mpc^{-1}}$, M$_B$ is the total B 
band luminosity of the galaxy from RC3, 
and B$_{TO}$ is the expected turnover in the 
globular cluster luminosity function. }\label{tab2}
\normalsize  
\end{table*} 

\begin{figure} 
\centerline{\psfig{figure=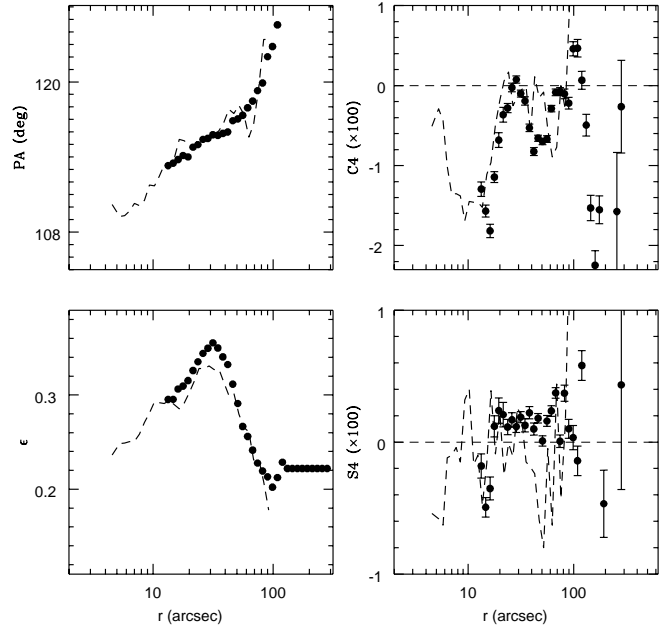,width=0.5\textwidth,angle=0}} 
 \caption
{Radial profiles of NGC 1052 in the $B$-band for the ellipticity
 ($\epsilon$), position angle (PA), 4th cosine term (C4)  and 4th sine
 term (S4), estimated by fitting ellipses to the galaxy light
 profile. At radii larger than $\approx130^{\prime\prime}$ both the
 ellipticity and the position angle are kept constant during the
 ellipse fitting routine, due to  S/N constraints. 
 Errors for the PA and $\epsilon$ are on the order of the symbol
 size. The dashed line
 shows the data of Peletier \etal (1990). The radius is the 
 major--axis radius.
}\label{n1052c4} 
\end{figure}

The elliptical galaxy NGC 1052 has been modelled by fitting ellipses
using the ISOPHOTE task within STSDAS (for the S0 galaxy 
NGC 7332 we used a median filter technique).  
During the ellipse fitting process  the centre of
the galaxy was kept fixed and a $3\sigma$ clipping algorithm was
employed. Bright objects, as well as the central saturated parts of 
the galaxy, were masked out. The position angle (PA) and ellipticity
($\epsilon$) were fit at each radius, until S/N constraints terminated
the fitting process at low surface brightness. 
The ellipticity, position angle, 
4th sine (S4) and 4th cosine (C4) Fourier components of the 
isophotal fits in the $B$-band are given in Figure \ref{n1052c4}.
Similar profiles and trends are seen in the $V$ and $I$ band images. 



\section{Globular cluster selection} \label{sec_sel}

To detect sources superimposed on the galaxies we first subtracted the
galaxy light from both NGC 1052 and 7332 
using a median filter and then followed the techniques 
described in Georgakakis et al. (2000). Briefly, sources are detected
using the  SExtractor package (version 2.1.0; Bertin \& Arnouts 1996)
with a  threshold of 2.5$\times \sigma_{sky}$. After detecting sources
in the three filters separately, they are matched to create a list of
common  detections. Magnitudes were calculated using 5$^{''}$
apertures plus small aperture  corrections to reach the total
derived magnitude.  

From the list of common detections, GC candidates are then selected on
the  basis of colour. In a similar way to Georgakakis  et al. (2000),
only those objects that fall within a region in  colour--colour  space
corresponding to Galactic GCs but extended to higher metallicity.  In
particular, objects must lie between $1.2<B-I<2.5$ and $0.6<V-I<1.7$
within $3\sigma$ photometric error.  Based on colour selection alone,
we find 513 and 215 GC candidates in NGC 1052 and NGC 7332
respectively.  After colour selection, we automatically removed
resolved objects based on their $V$-band FWHM size. A
final visual check was then conducted.

Completeness tests were carried out by adding artificial point sources
of different magnitudes to the original image, using MKOBJECTS task
which are then recovered using SExtractor. 
The $80\%$ completeness limits for NGC 1052 are found to be B $\approx
26.1$, V $\approx 25.6$ and  I $\approx 23.6$ mag. For NGC 7332 the
corresponding limits are  be B $\approx 26.8$, V $\approx 25.7$ and  I
$\approx 24.0$ mag. The mean photometric error at these magnitude
limits is about  $\approx0.1$\,mag for all three filters.  The galaxy
background does not significantly affect the completeness limits
beyond a galactocentric radius of $\sim65^{''}$.  
It should be noted that  including  GCs 
brighter than the $50\%$ completeness limit for point sources does not
change any of our conclusions.

\begin{figure} 
\centerline{\psfig{figure=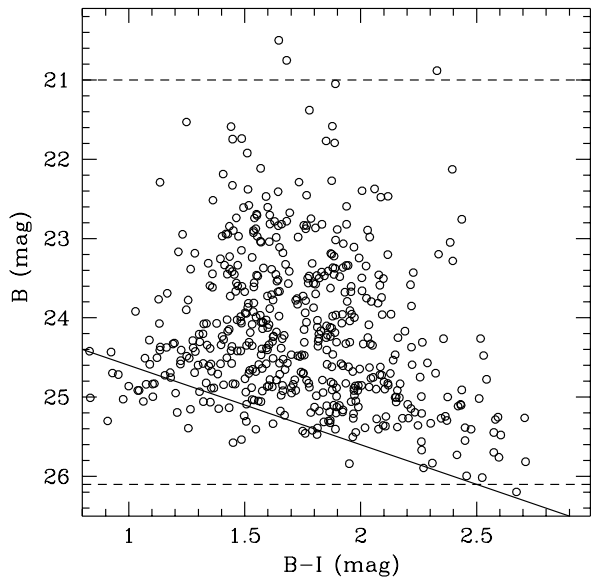,width=0.5\textwidth,angle=0}} 
 \caption
{Colour--magnitude diagram for globular clusters in NGC 1052, after
 colour selection and visual inspection.  
 The solid line corresponds to the completeness limit of
 $I=23.6$\,mag. The dashed lines are the $B$-band magnitude cutoffs
 for the selection of the final GC sample}
\label{n1052_cmd}     
\end{figure}

\begin{figure} 
\centerline{\psfig{figure=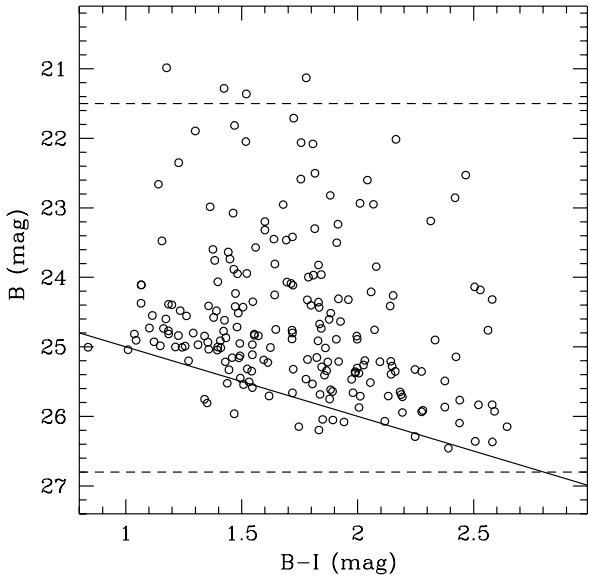,width=0.5\textwidth,angle=0}} 
 \caption
{Colour--magnitude diagram for globular clusters in NGC 7332, after
 colour selection and visual inspection.  
 The solid line corresponds to the completeness limit of
 $I=24.0$\,mag. The dashed lines are the $B$-band magnitude cutoffs
 for the selection of the final GC sample.}
\label{n7332_cmd}    
\end{figure}

The bright magnitude is set to be about 1 mag. brighter than
Omega Cen (the brightest GC in our Galaxy) at the distance of the
galaxy. Sources brighter than this are most likely stars.
Our magnitude limits are thus 21.0 $<$ B $< $ 26.1, 20 $<$ V $<$ 25.6,
19 $<$ I $<$ 23.6 for NGC 1052, and 21.5 $<$ B $<$ 26.8, 20.5 $<$
V $<$ 25.7, 19.5 $<$ I $<$ 24.0 for NGC 7332. 
After these various selection cuts 
we have 359 GC candidates in NGC 1052 and 154 in NGC 7332.
The colour--magnitude distribution of 
the samples is displayed in Figures \ref{n1052_cmd} and 
\ref{n7332_cmd}. Lines in these plots show the colour and
magnitude limits described above.

Although we attempt to minimise the fraction of foreground stars and
background galaxies in the final GC sample, it is likely that a number
of contaminating sources are also selected. Details of how we estimate
the contamination level in our Keck images can be found in
Georgakakis et al. (2000). Briefly, we used the Galaxy model
developed by Bahcall \& Soneira (1980) to 
estimate the contamination by foreground stars. 
After applying the same
magnitude and colour selection, the  Galaxy model predicts a total of
about $\approx$ 10 stars in the NGC 1052 GC sample and 24 in NGC
7332. Thus it both cases the number of contaminating  stars in the
final sample is negligible and will not effect our  conclusions.  

The number of background galaxies in our images are estimated from the 
galaxy count studies of Lilly, Cowie \& Gardner (1991). From the number
of predicted galaxies (within  our colour and magnitude range, and
field-of-view), we subtract the number of resolved objects suggesting
a total of about 70 
galaxies. Thus the contamination from relatively compact galaxies is
estimated to be 19\% for NGC 1052 and 45\% for NGC 7332. Simple
simulations in Georgakakis \etal (2000) suggest that these
contamination rates are an overestimate and the true rate is
somewhat less than this. 
We return to the issue of contamination in section 5.2.

\section{Results}\label{sec_res}

\subsection{NGC 1052}

\begin{figure} 
\centerline{\psfig{figure=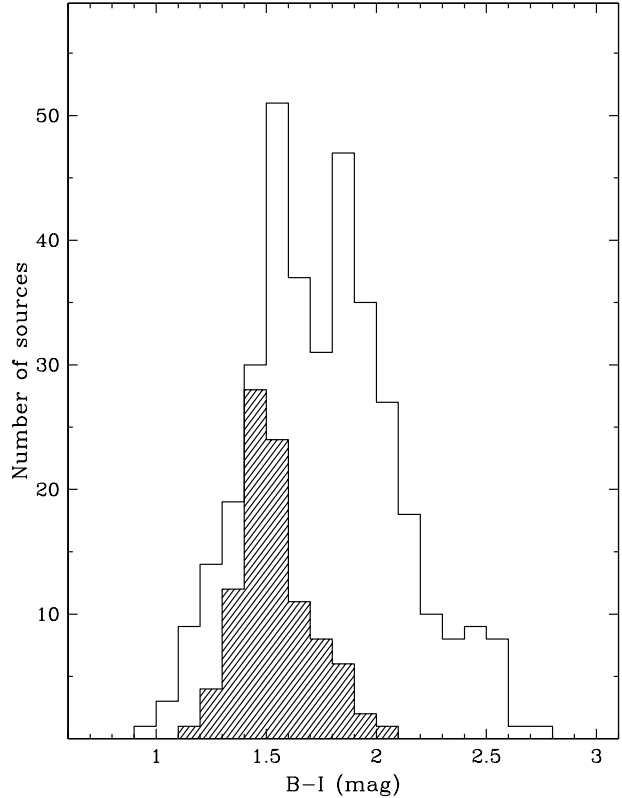,width=0.5\textwidth,angle=0}} 
 \caption
{Colour distribution for the globular clusters in NGC 1052. Bimodality
 in the B--I distribution is detected at the 99.7\% level from KMM
 statistics.  Also shown is the colour distribution of  Galactic
 globular clusters (hatched histogram).  
 The blue peak lies  close to the Milky Way
 globular cluster distribution (dominated by halo 
 clusters)}\label{n1052hist}          
\end{figure}

The colour distribution for the final NGC 1052 GC sample is plotted in
Figure \ref{n1052hist}. There is evidence for bimodality with peaks in
the distribution at $B-I\approx1.5$ and $B-I\approx1.9$. This is
confirmed by the KMM  statistical test (Ashman \etal 1994) which
rejects the single Gaussian model at a confidence level better than
$99.7\%$. Although the exact location of the peaks depend on which GCs
at the extremes of the distribution are excluded, the difference in
the peaks is nearly constant at around $\Delta$B--I = 0.4 $\pm$
0.1, where the error is estimated
assuming an estimated uncertainty of 0.07 mag in defining the peaks of the
bimodal distribution. 
Also shown
in  Figure \ref{n1052hist} is the reddening--corrected colour
distribution of  Galactic GCs.    
The blue peak of the NGC 1052 GCs is similar to that of the Milky Way
GC distribution, which is dominated by halo GCs. 
The KMM test finds roughly equal numbers of blue
and red GCs.

The contamination from background galaxies (i.e. 19\%) is unlikely to
significantly alter the results. In particular, the detection of
bimodality in the NGC 1052 GC system should be robust.  If we assume  
that the GCs comprise two distinct subpopulations of different age and
metallicity but with the same initial mass function, then we can
use the separation in the peak colours to constrain the age and
metallicity. We further assume that the blue (older) subpopulation has 
properties  similar to those of the Milky Way GC system and therefore
consists of old (15 Gyrs) metal--poor ($\mathrm{[Fe/H]=-1.5}$) GCs
(this is a reasonable assumption given the similarity in colour seen
in Figure  \ref{n1052hist} ). 
Here we employ the models of Worthey (1994) to predict the colour
difference between the red (younger) GC population and the old, 
metal--poor one. 

\begin{figure} 
\centerline{\psfig{figure=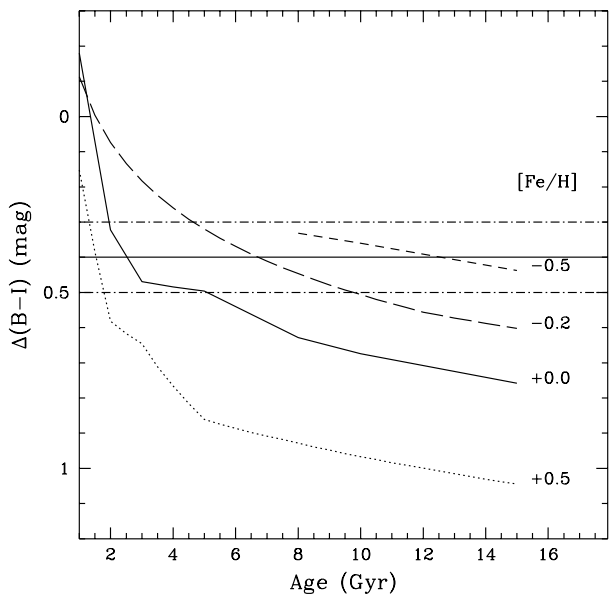,width=0.5\textwidth,angle=0}} 
 \caption
{Colour difference versus age diagram for NGC 1052 globular clusters
 using the Worthey (1994) models. Each curve corresponds to a
 different metallicity marked on the right of the curve. $\Delta(B-I)$
 is defined as the colour difference between the red and the blue
 (assumed 15 Gyrs old and metal--poor $\mathrm{[Fe/H]=-1.5}$) GC
 subpopulations.  The line corresponds to the measured colour
 difference in the  globular cluster peaks
 (i.e. $\Delta(B-I)=0.4\pm0.1$).
}\label{n1052model} 
\end{figure}

The model predictions for the B--I colour difference versus age of the
red subpopulation for different metallicities  are shown in Figure
\ref{n1052model}.  
The measured colour difference in the peaks from the KMM statistical
test is $\Delta(B-I)=0.4\pm0.1$,  where the error is estimated
assuming an uncertainty of 0.07 mag in defining the peaks of the
bimodal distribution. 

\begin{figure} 
\centerline{\psfig{figure=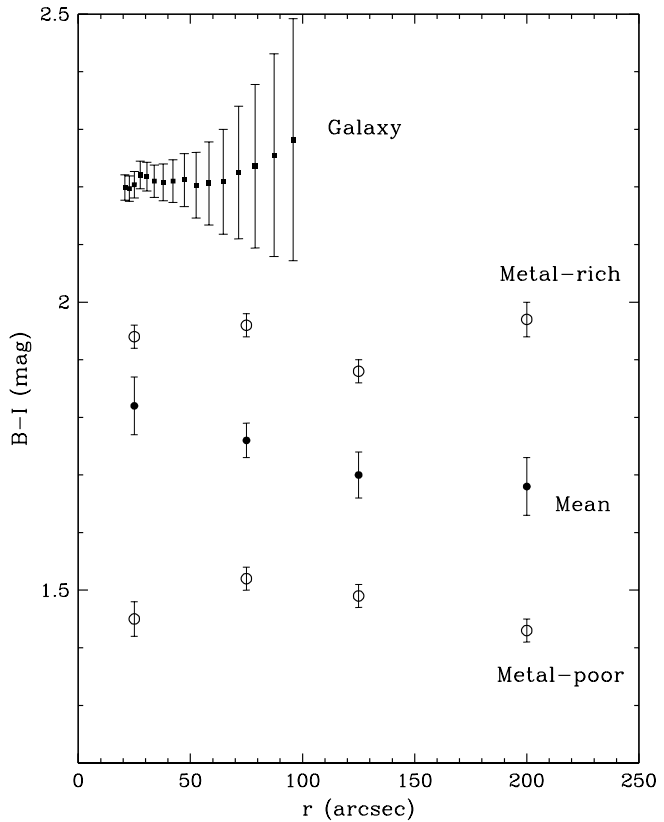,width=0.5\textwidth,angle=0}} 
 \caption
{Radial variation of B--I colour for the globular cluster 
subpopulations in NGC
1052. Filled circles show the mean for the whole system, and open
circles for the blue (metal--poor) and red (metal--rich)
subpopulations. The mean colour (metallicity) gradient is most
likely caused by a changing relative mix of the subpopulations
rather than an instrinsic colour gradient. The galaxy colour
profile is shown by filled squares. 
}\label{n1052rad} 
\end{figure}

In Figure \ref{n1052rad} we show the trend of GC colours with
galactocentric radius. The mean GC colour for the whole system
appears to getter bluer with increased distance from the galaxy
centre. However the mean colour of the blue (1.1 $<$ B--I $<$ 1.7)
and red (1.7 $<$ B--I $<$ 2.3) subpopulations are fairly constant
with radius. This suggests that the changing mean colour is due
to a changing relative mix of red and blue GCs with radius, rather
than any instrinsic radial gradient. The B--I profile for the
galaxy starlight is also shown in Fig. \ref{n1052rad}. It remains
fairly constant at B--I $\sim$ 2.2.

\begin{figure} 
\centerline{\psfig{figure=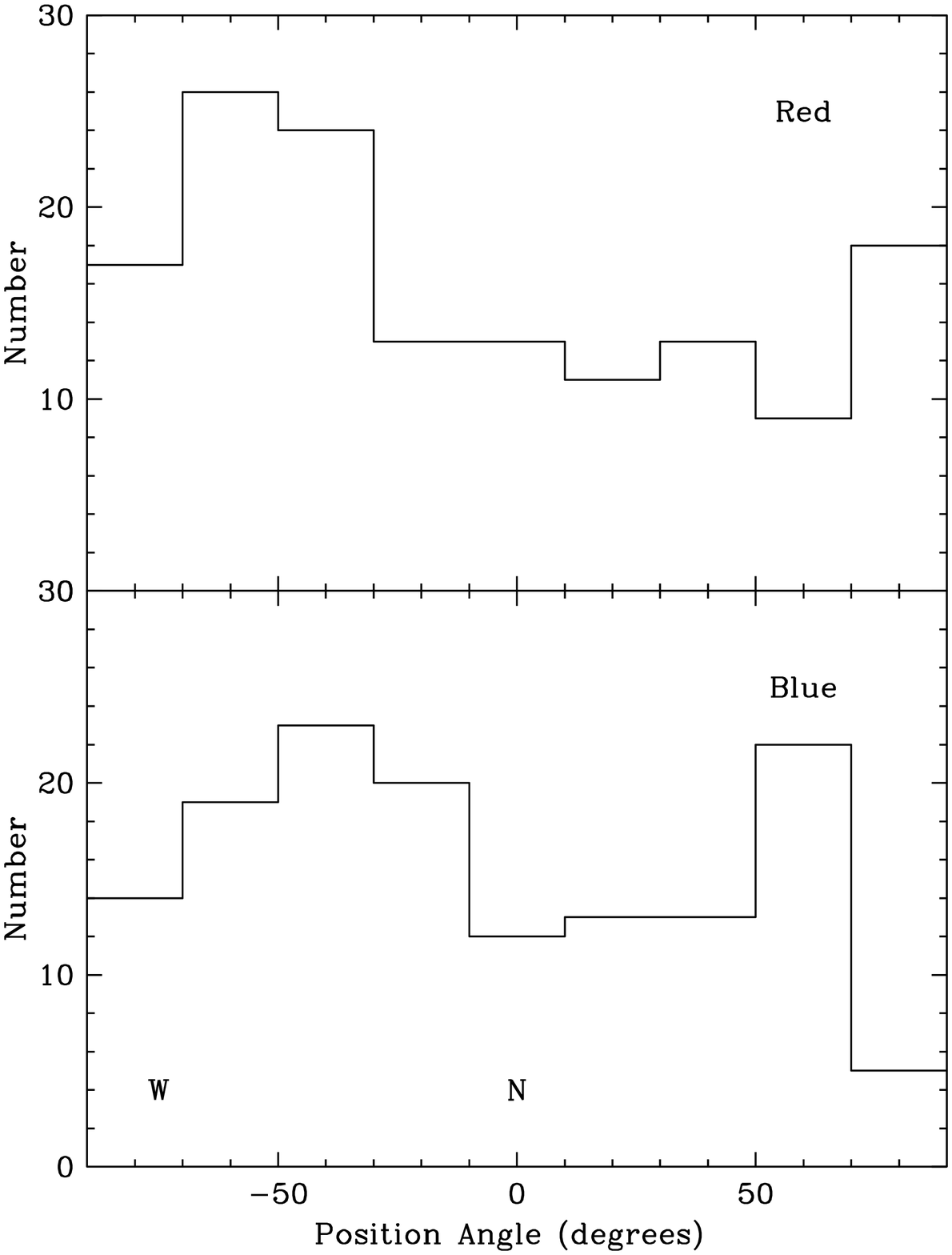,width=0.5\textwidth,angle=0}} 
 \caption
{Azimuthal distribution of globular clusters within 210$^{'}$ of
NGC 1052. The galaxy major axis is at position angle $\sim$ 120 or
--60$^{o}$. There is some evidence that the position angle of the 
red subpopulation is aligned with the galaxy major axis. 
}\label{n1052pa} 
\end{figure}

For the same blue and red subpopulations, we have examined their
azimuthal distribution. In Figure \ref{n1052pa} we show histograms of
position angle for GCs within 210 arcsec. Both subpopulations show
a slight tendency for the GCs to have an excess at a 
position angle similar to the galaxy major axis (and a deficit along the
galaxy minor axis). This effect is stronger for the red GCs.

\begin{figure*} 
\centerline{\psfig{figure=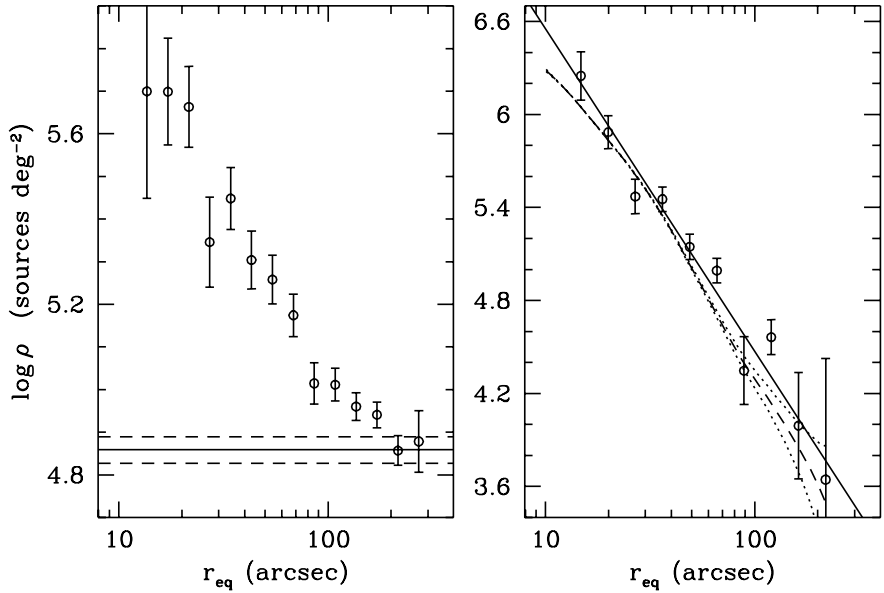,width=0.7\textwidth,angle=0}} 
 \caption
{{\bf (a)} Left panel: the surface density profile of all the objects  
 detected in the NGC 1052 V band frame. The solid line is the
 mean background surface density, estimated by combining the counts in the
 bins at large galactocentric radii. The dashed lines
 are the  errors around the mean assuming Poisson statistics. {\bf (b)}
 Right panel:  the surface density  profile of the globular cluster
 candidates, derived by statistically subtracting the
 background surface  density level and correcting for magnitude
 incompleteness.  The best fit  power law to the observed profile
 (solid line)  has the form  $\rho\propto r^{-2.08\pm0.13}$. The
 dashed line is the  surface brightness profile of NGC 1052 in
 arbitrary units. The dotted lines represent the $1\sigma$ surface
 brightness uncertainty. The globular cluster distribution is more
 extended  than the galaxy starlight. 
}\label{n1052dens}    
\end{figure*}

The surface density profile of GCs around NGC 1052 has been estimated
in the following way. From objects detected in 
the V band image, we selected point sources (based on their FWHM
and visual inspection) 
brighter than the $80\%$ completeness limit. Using
detections in a single band,  instead of the colour selected GC
sample, makes  the selection criteria easier to quantify. These
objects are then binned into elliptical rings centered on NGC 1052  and
having the same mean ellipticity (i.e. 0.2) and  position angle
(i.e. 115$^{\circ}$)  
as the galaxy. The counts in each bin are corrected for geometric 
incompleteness  due to foreground saturated stars, bad columns and the
limited size of the field--of--view (but we have not yet corrected
these counts for magnitude incompleteness). The surface density
profile corresponding to these counts in shown on the left of
Figure \ref{n1052dens}. It is clear that the counts decline from
the galaxy centre, confirming that the vast bulk of objects are
indeed associated with NGC 1052. 
After statistically subtracting the background from the number of
V band detected objects and after corrections for geometric 
and magnitude incompleteness (assuming V$_{TO}$ = 23.90 and
$\sigma$ = 1.15), we estimate 410 $\pm$ 45 globular
clusters in the LRIS frame. Integration of the surface density
profile (assuming a constant density at small radii) 
returns a similar number of GCs, i.e. 390. We adopt 400 as the
total number of GCs (i.e. covering all radii and magnitudes).  

By calculating the {\it incompleteness} in each radial bin 
and assuming a standard Milky Way GC luminosity
function, 
we correct the number counts. We also subtract a background
level, which at large radii is estimated to be  
$\rho_{bg}=7.23\pm0.05\times10^4\mathrm{\,sources\,deg^{-2}}$. 
The incompleteness-corrected and background-subtracted surface 
density profile is shown in the right side of Figure
\ref{n1052dens}. A fit to the profile indicates a slope of 
$\alpha=-2.08\pm0.13$. Earlier work by Harris \& Hanes (1985) found a 
GC  surface density slope of $\alpha=-2.26\pm0.27$. 
We adopt a total GC system for NGC 1052 of 400 $\pm$ 45, this
translates into a specific frequency of S$_N$ = 3.2 $\pm$ 0.37. 
In a previous
ground-based study, Harris \& Hanes (1985) estimated a total GC
population of 430 $\pm$ 80 for NGC 1052. 
Also shown on the right hand side of Figure \ref{n1052dens} is
the normalized surface brightness profile of NGC 1052. 
The GC density profile shows a similar decrease with radius as
the starlight. 

\begin{figure*} 
\centerline{\psfig{figure=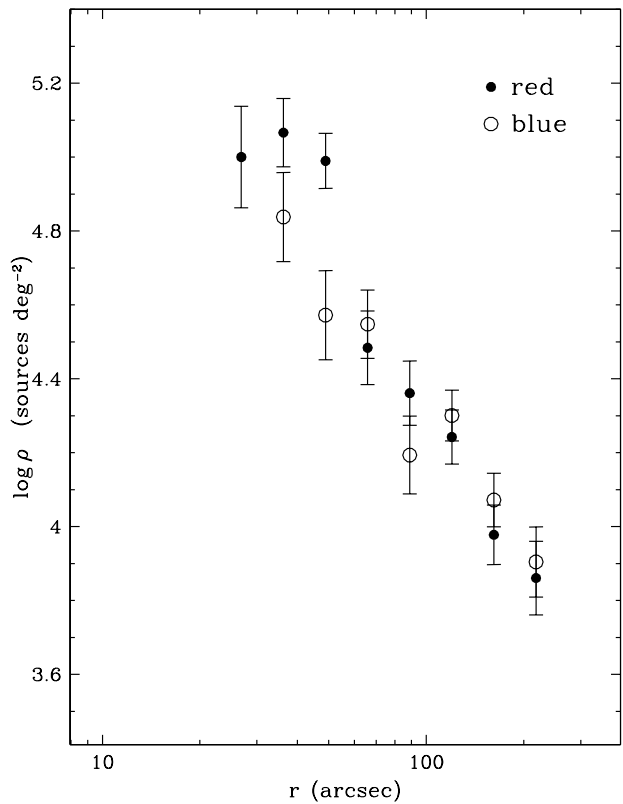,width=0.5\textwidth,angle=0}} 
 \caption
{ Globular cluster surface density of the red and blue
subpopulations. The red GCs are 
slightly more centrally concentrated than the
blue GCs. 
}\label{denrb}    
\end{figure*}

The density profile separated into 
red and blue GC subpopulations is shown in Figure \ref{denrb}. 
The red GCs are slightly more centrally concentrated than the
blue GCs.

\subsection{NGC 7332}

\begin{figure} 
\centerline{\psfig{figure=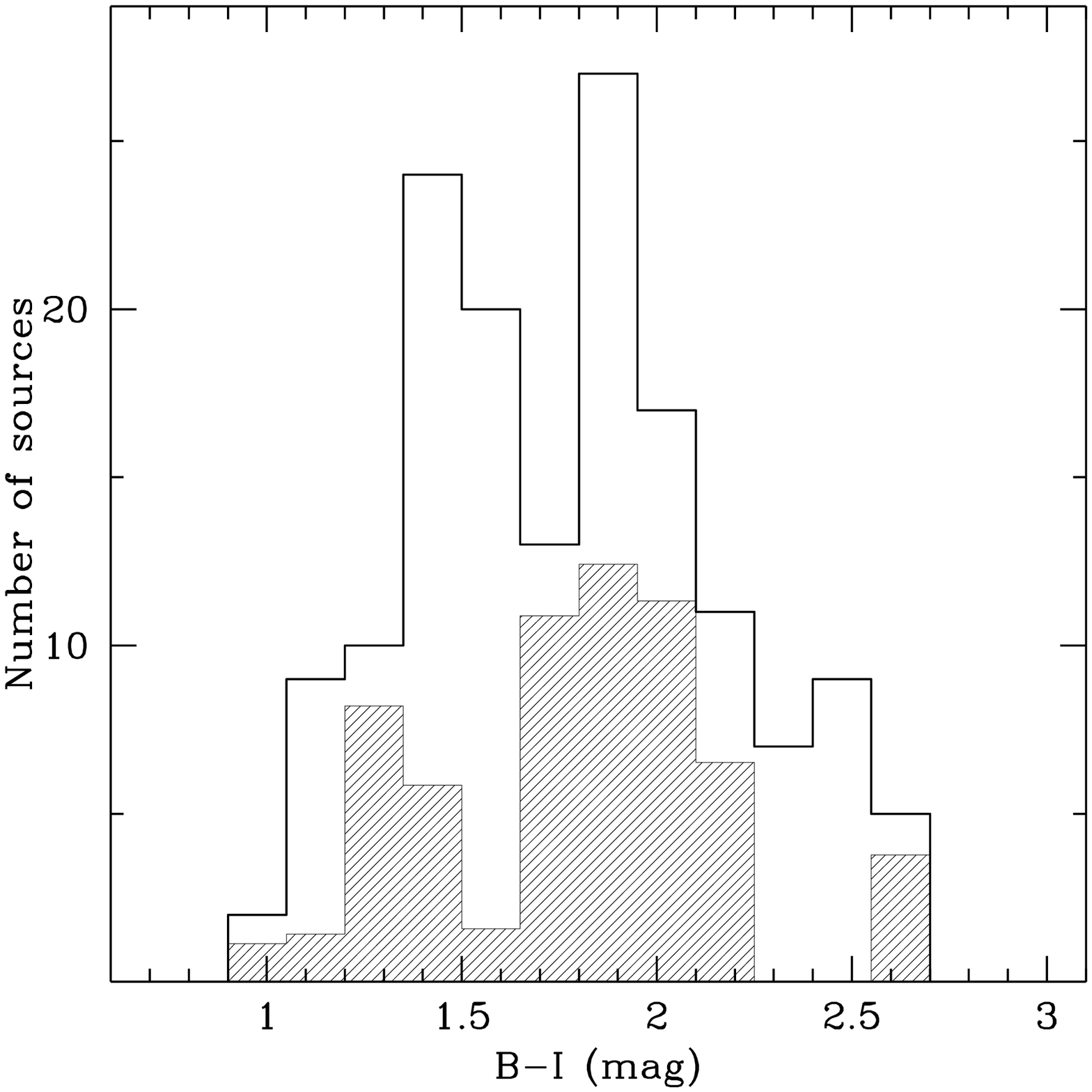,width=0.5\textwidth,angle=0}} 
 \caption
{Colour distribution for the globular clusters in NGC 7332, and
 predicted distribution of contaminating galaxies (hashed
 histogram). See text for details. 
}\label{n7332contam}         
\end{figure}

The colour distribution for the final NGC 7332 GC sample is plotted in
Figure \ref{n7332contam}. There is evidence for bimodality with the peaks
of the distribution at $B-I\approx1.45$ and
$B-I\approx1.95$. After rejecting a few of the reddest objects,
the KMM test rejects the single Gaussian model at
a confidence level better than $97.2\%$, and indicates a 
$\Delta$B--I = 0.5. Thus we have tentative
evidence for bimodality. 

Contamination by background galaxies, and to some extent
foreground stars, is more of an issue for NGC 7332. For the final
sample of 154 objects, we estimate that 45\% may be galaxies and
15\% stars. Figure \ref{n7332contam} also shows the expected  
colour distribution of galaxies, superposed on that for the GC
candidates. Foreground stars have a very broad colour distribution and
so  do not significantly modify the observed candidate GC colour
distribution.  Background galaxies, however, have similar colours. In
particular, a significant fraction of the red peak objects are
probably galaxies. 

\begin{figure} 
\centerline{\psfig{figure=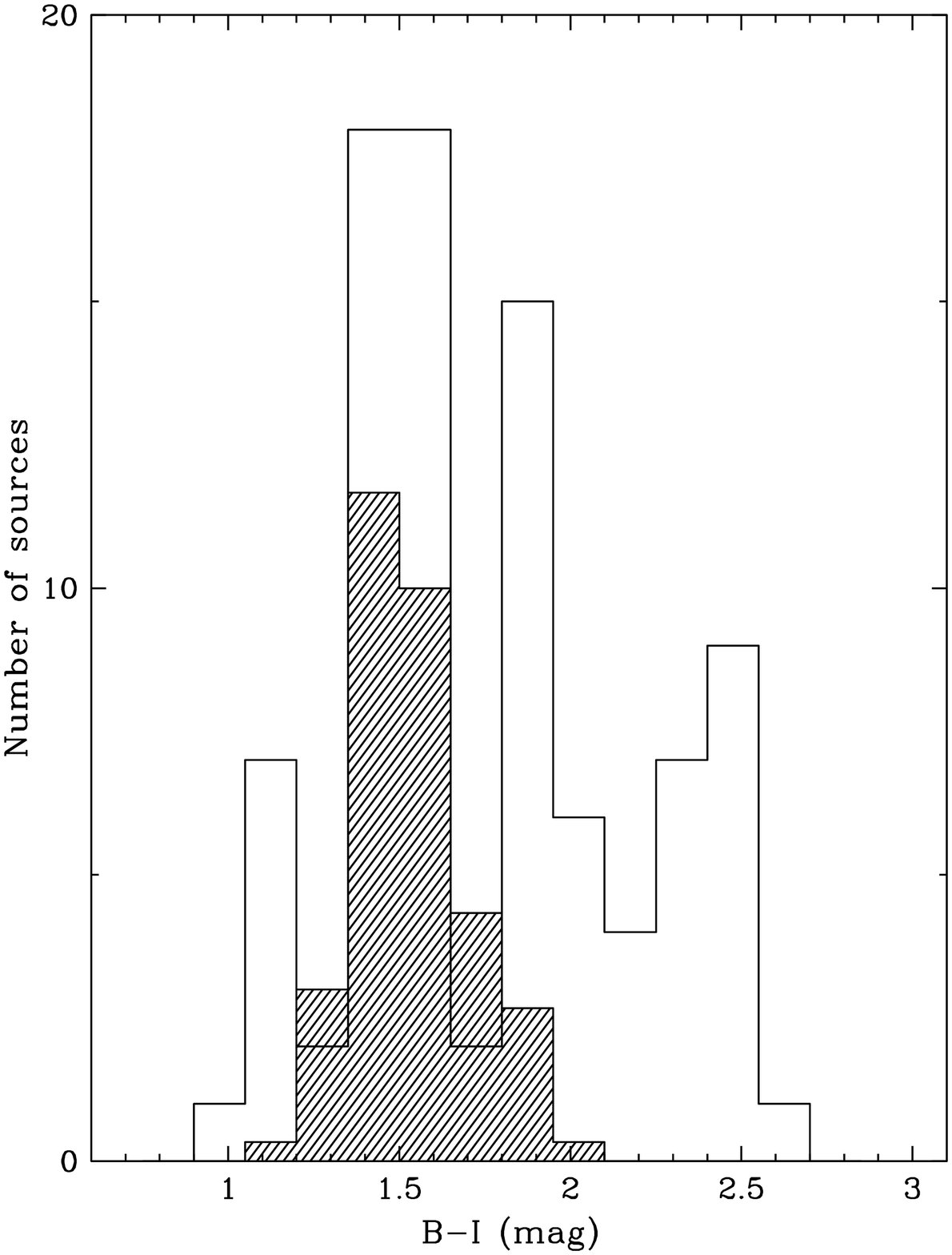,width=0.5\textwidth,angle=0}} 
 \caption
{Colour distribution for the globular clusters in NGC 7332 after
 subtraction of the estimated contamination from background
 galaxies. There is
 tentative evidence for bimodality in the B--I distribution from KMM
 statistics.  Also shown is the colour distribution of Galatic
 globular clusters  
 (hatched histograms).  The blue peak lies close to the 
 Milky Way distribution (which is dominated by halo globular 
 clusters)}\label{n7332hist}          
\end{figure}

We subtract the galaxy colour distribution, and show the residual
GC colour distribution in Figure \ref{n7332hist}. A narrow peak at 
B--I $\sim$ 1.5 remains, while the red distribution becomes much
broader and less well defined. We also show in Figure
\ref{n7332hist} is the 
reddening--corrected colour 
distribution of Galactic GCs.  
Like NGC 1052, the blue peak of the NGC 7332 GCs is similar to that 
of the Milky Way GC distribution (which is dominated by halo
GCs).

Given the similarity to the Milky Way colour distribution and the
small number of blue galaxies, the reality of the blue GC peak
seems fairly secure. However it is more difficult to be confident
about the red peak. There are too few sources to investigate the
red ones separately with any confidence. We proceed assuming that
NGC 7332 does indeed have two GC subpopulations but any results
should be regarded as tentative at best.
 
\begin{figure} 
\centerline{\psfig{figure=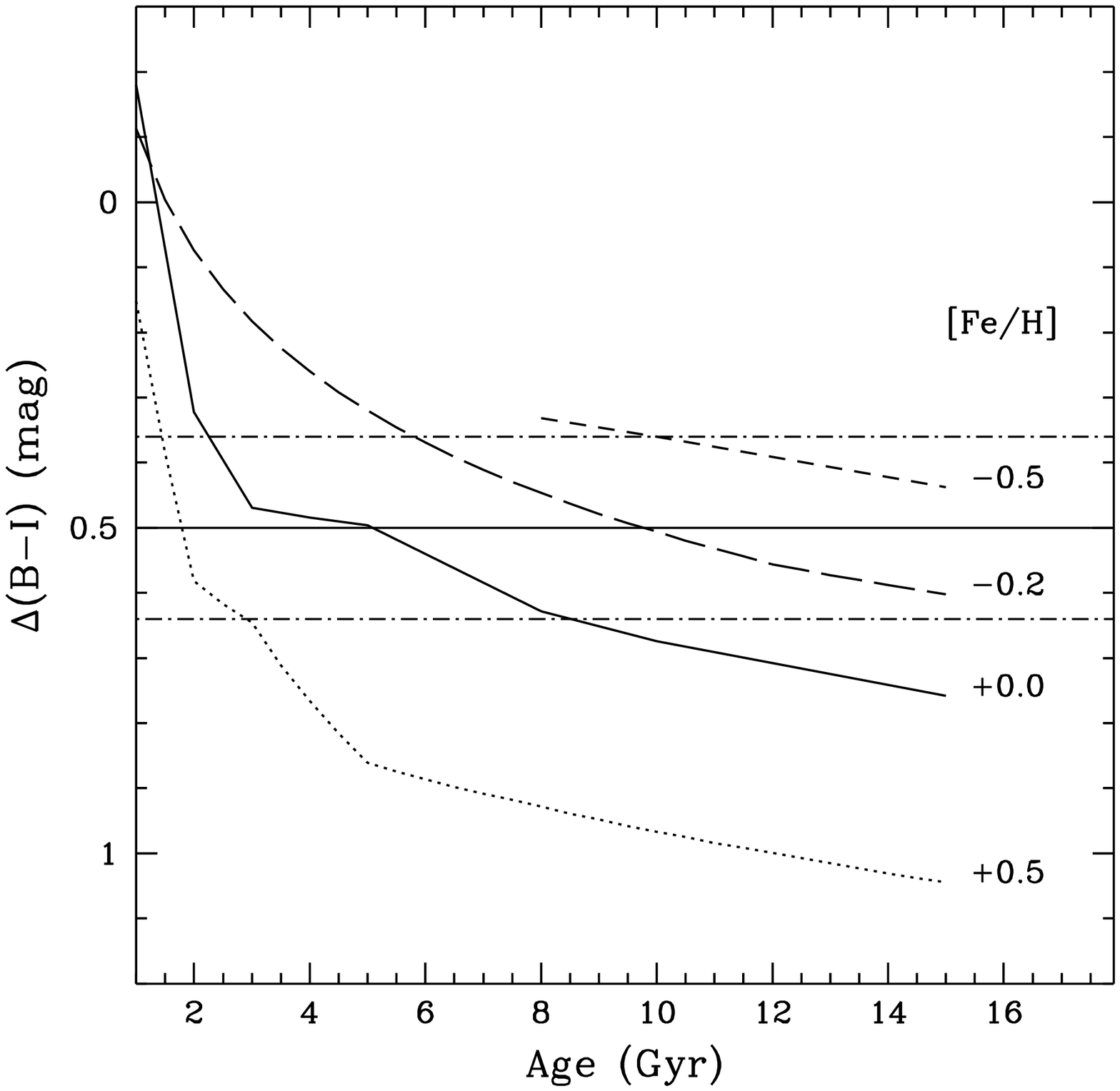,width=0.5\textwidth,angle=0}} 
 \caption
{Colour difference versus age diagram for NGC 7332 globular clusters
 using the Worthey (1994) models. Each curve corresponds to a
 different metallicity marked on the right of the curve.
 $\Delta(B-I)$ is defined as the colour difference between the red and
 the blue (assumed to be 15 Gyr old and metal--poor
 $\mathrm{[Fe/H]=-1.5}$) GC subpopulations.  The line corresponds to
 the measured colour difference in the  globular cluster peaks
 (i.e. $\Delta(B-I)=0.50\pm0.14$). 
}\label{n7332model} 
\end{figure}

Following the same assumptions as above for NGC 1052, we can
investigate the age and metallicity of the possible two GC
subpopulations in NGC 7332.  Again, relative to old  (15 Gyrs),
metal--poor ($\mathrm{[Fe/H]=-1.5}$) GCs, the age of the red GCs can
be constrained from the measured colour difference of
$\Delta(B-I)=0.5\pm0.14$ (we have assumed $\pm$0.1 for the individual
peak colours).  This color difference and the models of Worthey (1994)
are shown in Figure \ref{n7332model}.

\begin{figure*} 
\centerline{\psfig{figure=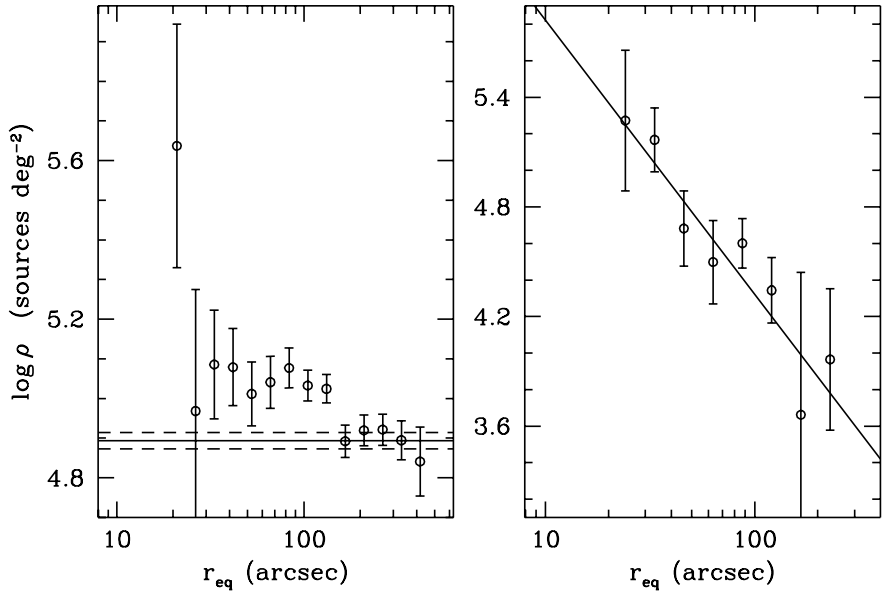,width=0.7\textwidth,angle=0}} 
 \caption
{{\bf (a)} Left panel: the surface density profile of all the objects 
 detected in the NGC 7332 V band frame. The continuous lines is the
 mean background surface density, estimated by combining the counts in the
 bins at large galactocentric radii. The dashed lines
 are the  errors around the mean assuming Poisson statistics. {\bf (b)}
 Right panel:  the surface density  profile of the globular cluster
 candidates, derived by statistically subtracting the
 background surface  density level and correcting for magnitude
 incompleteness. The best fit  power law to the observed profile
 (continuous line) has the form  $\rho\propto r^{-1.5\pm0.23}$. 
}\label{n7332dens}    
\end{figure*}

In Figure 14 
we show the surface density distribution of GC
candidates around NGC 7332. This has been carried out in a
similar way to NGC 1052, i.e. using the V band image we correct
the counts in elliptical bins for geometric incompleteness and
the limited field-of-view. These counts are converted into
surface density and shown in the left panel of Figure 14. 
After background subtraction and correction for
magnitude incompleteness, we derive the profile shown in the
right panel. A fit to the final surface density profile indicates
a slope $\alpha$ = --1.50 $\pm$ 0.23. 
The background level is estimated at
$7.85\pm0.37\times10^{4}\,\mathrm{\,sources\,deg^{-2}}$. 
Subtracting this from the V band detections (after 
corrections for geometry, magnitude incompleteness etc) 
leaves 164 $\pm$ 35 objects.  
Integration of the surface density
profile (assuming a constant density at small radii) 
returns a slightly higher estimate of 219 GCs. 
If we adopt 190 $\pm$ 30 as the total GC system, then it implies 
a specific frequency of S$_N$ = 2.0 $\pm$ 0.3. 

We attempted to separate the density profile into the 
blue and red subpopulations. They show little difference, but do
confirm that both subpopulations decline in density from the
galaxy centre at least out to 100 arcsec. This suggests that
some fraction of {\it both} the blue and red subpopulations are
indeed associated with the galaxy. This gives us some extra
confidence in the reality of the GC colour bimodality.

\section{Discussion}\label{sec_disc}

\subsection{NGC 1052}

Located in a small group, this E4 galaxy contains an active
nucleus. Interesting features include a LINER spectrum and variable
radio core (Fosbury \etal 1978), luminous H$_2$O maser (Braatz, Wilson
\& Henkel 1994) and an absorbed X--ray emitting nuclear source (Weaver
\etal 1999).  Although the outer stellar envelope 
appears to be relatively undisturbed
(it has a relatively low fine structure value of $\Sigma$ = 1.78;
Schweizer \& Seitzer 1992), it reveals multiple evidence for a 
past merger: \\

\noindent
$\bullet$ Infalling HI gas. HI is seen in absorption redshifted on
the nucleus indicating infalling HI gas (van Gorkom \etal
1986).\\
$\bullet$ HI tidal tails. Extended HI mapping reveals two tidal
tails, indicative of a gaseous merger about 1 Gyr ago (van Gorkom
\etal 1986). \\
$\bullet$ Misaligned star/gas axis. The rotation axis of the HI
gas (PA = 134$^o$) and ionized gas (125-131$^o$) is misaligned with
the stellar rotation axis (28$^o$). 
This indicates an external origin for the gas (van Gorkom
\etal 1986). \\
$\bullet$ Dust lanes. Optical imaging (Forbes, Sparks \&
Macchetto 1990) indicates the presence of dust lanes near the
galaxy centre. Such a structure is unlikely to arise from {\it in
situ} stellar mass loss. \\

Evidence for a past merger/accretion event seems
overwhelming. However the time since the merger warrants further
study. The HI tails would suggest a very recent merger
(i.e. $\sim$ 1 Gyr), whereas the lack of optical disturbance and
a Fundamental Plane residual of +0.07 (Prugniel \& Simien 1996) 
would suggest a long time has elapsed since the last
major burst of star formation (Forbes, Ponman \& Brown
1998). A reliable spectroscopic age for the galaxy stars 
is not yet available. 
To be consistent with the time since the HI tails formed, the red
GCs would need to have a metallicity of at least solar (for 
[Fe/H] $\ge$ 0, the implied age from Figure \ref{n1052model} is
$\le$ 2.5 Gyrs). If the red GCs were indeed this metal--rich and
young, it would support arguments for 
their formation in the same
merger event that formed the HI tails. However, GCs this young
would also be expected to be brighter on average than the blue
GCs, even if metal--rich (Worthey 1994). Examination of
Figure 4 shows that this is not the case, i.e. 
the red GCs are about 0.5 mag {\it
fainter} on average than the blue ones. The current HI content, of a few
10$^{8}$ M$_{\odot}$ (van Gorkom \etal 1986),  
suggests that the most recent merger will not produce 
significant numbers of GCs for typical gas-to-cluster formation
efficiencies. 
So although a recent merger has occurred in NGC 1052, that event seems
very unlikely to be responsible for the population of red GCs. 
Globular cluster formation in a 
very old merger, via multi--phase collapse or from accretion would
appear more likely indicating that the 
red GCs are metal--poor, e.g. [Fe/H] $\sim$ --0.5. 

The properties of NGC 1052 GC system appear to be unexceptional. We
derive a specific frequency, S$_N$, of 3.2. This is typical for
ellipticals in groups (Harris 1991).  
With S$_N$ = 3.2 and a GC density profile slope of $\alpha$ =
--2.08, the galaxy lies is within the S$_N$--slope trend seen by Forbes
\etal (1997). 
Like other galaxies (Geisler \etal 1996; Forbes \etal 1998) the
apparent radial gradient in GC mean colour appears to be due to
the fact that the red (metal--rich) GCs are more centrally
concentrated than the blue (metal--poor) ones. Thus the 
intrinsic gradient is weak. In a collapse scenario this
suggests that dissipation has not played
a strong role. The galaxy starlight has a similar (it is redder
by about 0.25 mags) colour than the red GCs, as is often seen in other
galaxies (e.g. Forbes \& Forte 2000). It suggests that the red
GCs and the galaxy field stars may have formed at similar times
from similar enriched gas. 
The red, and to a lesser extent the blue, GC systems appear to be
aligned in position angle with the galaxy major axis. 

In terms of GC formation models, most of the GC system properties
would be consistent with those expected from either an old merger
(Ashman \& Zepf 1992) or the multi--phase collapse (Forbes \etal 1997).
One feature that favours the multi--phase collapse idea is the
tendency for the red GCs to be more closely aligned in position
angle to the galaxy
starlight than the blue GCs. This is a direct consequence of the
red subpopulation forming during the `galactic' phase in which
the bulk of the galaxy stars also form. 

\subsection{NGC 7332}

This S0 galaxy reveals evidence for distortions in the outer
isophotes. Schweizer \& Seitzer (1992) assigned it 
$\Sigma$ = 4.00, indicating high optical disturbance. It also
lies off the Fundamental Plane by --0.31 dex (Prugniel \& Simien
1996) suggesting a young age (Forbes \etal 1998). Indeed
spectroscopy of the galaxy centre indicates an age since the last
major burst of star formation of 4.5 Gyrs (Terlevich \& Forbes
2000). Studies of the ionized
gas reveal evidence for two components, one of which is
counter--rotating with respect to the stellar distribution (Plana
\& Boulesteix 1996). Such evidence suggests that NGC 7332 is a
good candidate for a recent merger event. 

If the red GCs formed in a merger event less than 5 Gyrs ago,
then a metallicity of at least solar is indicated by 
the Worthey (1994) models (see Figure 13). This
would be consistent with the spectroscopic age of the galaxy, but we
remind the reader of the tentative nature of the bimodal colour
distribution. 
An old age for the red GCs (expected under the accretion and
multi--phase collapse models) would require they be metal--poor,
e.g. around half solar. 
Assuming that the red subpopulation is indeed
dominated by GCs, we can not currently discriminate between GC
formation in a recent merger event or {\it in situ} associated
with the bulge/disk of the galaxy. 

We derive a relatively shallow 
slope for the radial GC surface density of 
$\alpha$ of --1.5, and a low specific frequency S$_N$ of
2.0, however such values remain uncertain due to the possible
high contamination rate.

\section{Conclusions}\label{sec_conc}

Here we present multi--colour Keck imaging of two nearby galaxies
thought to have undergone a merger. After colour, magnitude and
visual selection we detect 359 and 154 globular cluster
candidates in NGC 1052
and NGC 7332 respectively. 
We derive globular 
cluster density profiles, global specific frequencies and in
the case of NGC 1052, globular cluster radial colour gradients 
and azimuthal
distribution. In general these globular cluster properties are normal for
early--type galaxies. 

Both galaxies also reveal a bimodal globular cluster 
colour distribution indicating two globular 
cluster subpopulations (although in
the case of NGC 7332 the red subpopulation is heavily
contaminated by background galaxies). The KMM test indicates
roughly equal numbers in each subpopulation. The blue
subpopulation has colours consistent with being very old (age
$\sim$ 15 Gyrs) 
and metal--poor ([Fe/H] $\sim$ --1.5). 
The mean colour
difference in the two subpopulations is compared to stellar
population models. 
Although both galaxies have undergone a recent merger, we argue
that this event has had little effect on the GC systems. 
If globular clusters formed in a
multi--phase collapse or via accretion, the red ones are required to
be metal--poor ([Fe/H] $\sim$ --0.5).  
The key to
discriminating between the models is to determine the age
and/or 
metallicity of individual globular clusters with high S/N spectroscopy.

\section{Acknowledgements}\label{sec_ack}

We thank Soeren Larsen for useful comments and
suggestions. We also thank the referee Bill Harris for his
insightful comments. 
Part of this research was funded by NATO Collaborative
Research grant CRG 971552 and NSF grant AST 9900732. 
The data presented herein were obtained at the
W.M. Keck Observatory, which is operated as a scientific partnership among
the California Institute of Technology, the University of California and
the National Aeronautics and Space Administration.  The Observatory was
made possible by the generous financial support of the W.M. Keck
Foundation. This research has made use of the NASA/IPAC Extragalactic
Database  (NED), which is operated by the Jet Propulsion Laboratory,
Caltech, under contract with the National Aeronautics and Space
Administration.


\begin{thebibliography}{} 

{\bibitem{1} Ashman, K. A., Bird, C. M., Zepf, S. E.,
1994, AJ, 108, 2348}


{\bibitem{3} Ashman, K. M., Zepf S. E., 1992, ApJ, 384, 50}

{\bibitem{4} Bahcall, J. N., Soneira, R. M., 1980, ApJS, 44, 73}



{\bibitem{7} Bertin, E., Arnouts, S., 1996, A\&AS, 117, 393}



{\bibitem{98} Braatz, J. A., Wilson, A. S., Henkel, C., 1994, ApJ,
437, L99} 


{\bibitem{11} Brown, R. J. N., Forbes D. A., Kissler-Patig M., Brodie J.,
2000, MNRAS, 317, 406}


{\bibitem{13} Cote, P., Marzke, R. O., West, M. J., 1998, ApJ,
501, 554}



{\bibitem{96} Forbes, D.A., Sparks, W., Macchetto, F.D., 1990, In
{\it Paired and Interacting Galaxies}, p 431, ed J. Sulentic,
W. Keel, C. Telesco, NASA.}


{\bibitem{21} Forbes, D. A., Ponman, T. J., Brown, R. J. N.,
1998, ApJ, 508, L43}

{\bibitem{22} Forbes, D. A., Brodie, J. P., Grillmair, C. J.,
1997, AJ, 113, 1652} 

{\bibitem{23} Forbes, D. A., Grillmair, C. J., Williger, G. M.,
Elson, R. A. W., Brodie, J. P., 1998, MNRAS, 293, 325}


{\bibitem{97} Fosbury, R.A.E., Melbold, U., Goss, W.M., Dopita,
M.A., 1978, MNRAS, 183, 549}

{\bibitem{24} Georgakakis, A., Forbes, D. A., Brodie, J. P.,
2000, MNRAS, in press}

{\bibitem{25} Geisler, D., Lee, M. G., Kim, E., 1996, AJ, 111, 1529}


{\bibitem{28} Harris, W. E., 1991, ARA\&A, 29, 543}

{\bibitem{29} Harris, W. E., Hanes, D. A., 1985, ApJ, 291, 147}

{\bibitem{35} Landolt, A. U., 1992, PASP, 104, 336}

{\bibitem{37} Lilly, S. J., Cowie, L. L., Gardner, J. P., 
1991, ApJ, 369, 79}



{\bibitem{45} Oke, J. B., Cohen, J. G., Carr, M., Cromer, J., Dingizian, A.,
Harris, F., H. Labrecque, S.,  Lucinio, R., Schaal, W., Epps, 
H., Miller, J., 1995, PASP, 107, 375}  

{\bibitem{47} Plana, H., Boulesteix, J., 1996, A\&A, 307, 391}

{\bibitem{46} Prugniel, P., Simien, F., 1996, A\&A, 309, 749}

{\bibitem{99} Puzia, T. H., Kissler--Patig, M., Brodie, J. P.,
Huchra, J. P., 1999, AJ, 118, 2734}

{\bibitem{88} Schlegel, D. J., Finkbeiner, D. P., Davis, M.,
1998, ApJ, 500, 525}

{\bibitem{48} Schweizer, F., 1987, in Nearly Normal Galaxies, ed. S. Faber
(Springer: New York), 18}


{\bibitem{50} Schweizer, F., Seitzer, P., 1992, AJ, 104, 1039}


{\bibitem{52} Terlevich, A. I., Forbes, D. A., 2000, MNRAS, submitted}

{\bibitem{53} van Gorkom, J. H., \etal 1986, AJ, 91, 791} 

{\bibitem{61} Worthey, G., 1994, ApJS, 95, 107}

{\bibitem{62} Whitmore, B. C., Miller, B. W.,
 Schweizer, F., Fall, S. M., 1997, AJ, 114. 1797}



\end{thebibliography}
\end{document}